\documentclass[aps,twocolumn,superscriptaddress]{revtex4-2}
\usepackage[colorlinks=true,bookmarks=true,citecolor=magenta,urlcolor=magenta,linkcolor=magenta,breaklinks]{hyperref} 
\usepackage{breakurl}
\usepackage{bbm}
\usepackage{dsfont}
\usepackage{sidecap}
\usepackage{amssymb}
\usepackage{hhline}
\usepackage{urwchancal}
\usepackage{xcolor}
\sidecaptionvpos{figure}{t}
\usepackage{amsmath}
\usepackage{graphicx}
\usepackage{esint}
\usepackage{epstopdf}
\usepackage{rotating}
\graphicspath{{pict/}{./}}
\usepackage{bm}%

\newcommand\pictc[5]{\begin{figure}
                       \centerline{\vspace{-1mm}
\includegraphics[width=#1\columnwidth,height=0.7\textheight,keepaspectratio]{#3}}
                       \protect\caption{\protect\label{#4} #5}\vspace{-3mm}
                    \end{figure}            }

\newcommand\pict[4][1]{\pictc{#1}{!tb}{#2}{#3}{#4}}
\newcommand\rpict[1]{\ref{#1}}

\newcounter{Fig}

\newcommand\mymapstol{\mathrel{\ooalign{$\leftarrow$\cr%
  \kern1.75ex\raise0.275ex\hbox{\scalebox{1}[0.4]{$\mid$}}\cr}}}

\newcommand\mymapstor{\mathrel{\ooalign{$\rightarrow$\cr%
  \kern-.15ex\raise.275ex\hbox{\scalebox{1}[0.4]{$\mid$}}\cr}}}
\begin{document}


\title{Fabry-Pérot Metacavities with Single-Layered Dielectric Metamirrors}
\author{Zhichun Qi}
\email{Authors contribute equally to this work.}
\affiliation{College for Advanced Interdisciplinary Studies and Nanhu Laser Laboratory, National University of Defense Technology, Changsha 410073, P. R. China.}
\author{Chunchao Wen}
\email{Authors contribute equally to this work.}
\affiliation{College for Advanced Interdisciplinary Studies and Nanhu Laser Laboratory, National University of Defense Technology, Changsha 410073, P. R. China.}
\affiliation{Hunan Provincial Key Laboratory of Novel Nano-Optoelectronic Information Materials and Devices, National University of Defense Technology, Changsha, Hunan 410073, P. R. China.}
\author{Wei Wang}
\affiliation{College for Advanced Interdisciplinary Studies and Nanhu Laser Laboratory, National University of Defense Technology, Changsha 410073, P. R. China.}
\author{Jianhua Shi}
\affiliation{College for Advanced Interdisciplinary Studies and Nanhu Laser Laboratory, National University of Defense Technology, Changsha 410073, P. R. China.}
\author{Chucai Guo}
\email{Email: gcc\_1981@163.com}
\affiliation{College for Advanced Interdisciplinary Studies and Nanhu Laser Laboratory, National University of Defense Technology, Changsha 410073, P. R. China.}
\affiliation{Hunan Provincial Key Laboratory of Novel Nano-Optoelectronic Information Materials and Devices, National University of Defense Technology, Changsha, Hunan 410073, P. R. China.}
\author{Wei Liu}
\email{Email: wei.liu.pku@gmail.com}
\affiliation{College for Advanced Interdisciplinary Studies and Nanhu Laser Laboratory, National University of Defense Technology, Changsha 410073, P. R. China.}
\affiliation{Hunan Provincial Key Laboratory of Novel Nano-Optoelectronic Information Materials and Devices, National University of Defense Technology, Changsha, Hunan 410073, P. R. China.}

\begin{abstract}
The Fabry-P\'{e}rot resonator is a cornerstone of photonics and wave physics, providing a universal mechanism for spectral confinement and resonant enhancement of wave-matter interactions. In this work, we establish an analytically tractable class of Fabry-P\'{e}rot metacavities in which the reflecting elements are realized by single-layer periodic arrays of circular dielectric cylinders acting as metamirrors. Both the reflection efficiency and reflection phase of such metamirrors are obtained in closed form and shown to be widely and independently tunable, encompassing ideal electric and magnetic mirror limits with unit reflectivity. Building on these results, we derive explicit analytical expressions that fully describe the optical responses of Fabry-P\'{e}rot cavities composed of two such parallel metamirrors. Our combined analytical and numerical investigations reveal that these metamirrors provide exceptional flexibility for tailoring Fabry-P\'{e}rot resonances across a broad spectral range, enabling precise control over resonance positions and quality factors. In particular, the framework naturally predicts the emergence of Fabry-P\'{e}rot bound states in the continuum with formally infinite Q-factors. These results establish dielectric-metamirror-based Fabry-P\'{e}rot cavities as a versatile and fundamentally transparent platform for engineering high-Q optical resonances. 
\end{abstract}

\maketitle

\section{Introduction}

The Fabry-P\'{e}rot resonator composed of two parallel reflecting mirrors is a foundational structure for the confinement and spectral control of not only electromagnetic waves, but also physical waves of other forms~\cite{BORN1999a,SIEGMAN_lasers_1986,YARIV_2006_Photonics,DATTA_2009__Electronic,DUNN_2015_Springer}, forming the basis for a wide range of applications relying on Fabry-P\'{e}rot resonances, including the first detection of gravitational waves~\cite{JARAMILLO_2022_Phys.Rev.Lett._Gravitational}. The central optical elements of reflecting mirrors could be interfaces of high index contrast, metallic mirrors, Bragg mirrors of various forms and dimensions, and mirrors consisting of random media~\cite{BORN1999a,SIEGMAN_lasers_1986,YARIV_2006_Photonics,Joannopoulos2008_book,OTHONOS_1999__Fiber,segev2013anderson}. In the past decade, largely stimulated by the field of metasurfaces~\cite{YU_NatMater_flat_2014,CHEN_Rep.Prog.Phys._review_2016} and Mie-tronics~\cite{LIU_2018_Opt.Express_Generalized,KIVSHAR_2022_NanoLett._Rise}, new types of reflecting metamirrors consisting of high index particles are widely employed, the optical functionalities of which rely on resonant excitations of Mie resonances and their far-field interferences~\cite{jahani_alldielectric_2016,KUZNETSOV_Science_optically_2016,LIU_2018_Opt.Express_Generalized,KIVSHAR_2022_NanoLett._Rise,BABICHEVA_2024_Adv.Opt.Photon.AOP_Mieresonanta}. 
For such metamirrors, the reflection efficiency (even a single-layered metamirror can render perfect unit reflection) and reflection phase can be freely tuned~\cite{LIU_2018_Opt.Express_Generalized,liu_generalized_2017},  which are two central factors to exploit for efficient manipulations of  Fabry-P\'{e}rot resonances and their applications~\cite{FEIS_2020_Phys.Rev.Lett._HelicityPreserving,BASSLER_2024_Phys.Rev.Lett._MetasurfaceBased,ALAGAPPAN_2024_Phys.Rev.Lett._FabryPerot}.

\pict[1]{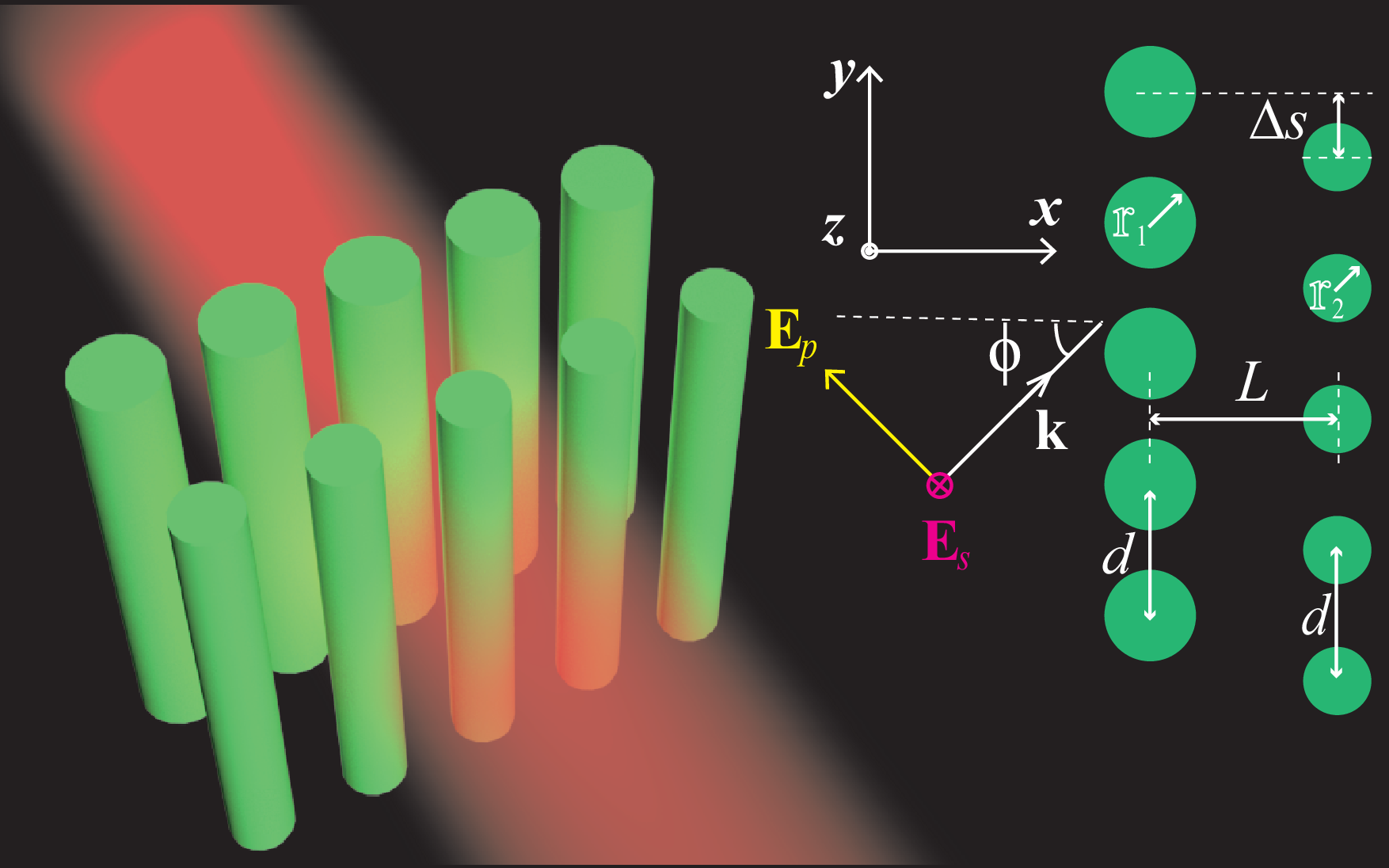}{fig1}{\small Schematic illustration of a plane wave (wavevector $\textbf{k}$ and incident angle $\phi$) interacting with a Fabry-P\'{e}rot metacavity consisting of two parallel periodic arrays of dielectric cylinders (metamirrors): the two arrays share the same period $d$ and cylinder refractive index $\mathbbm{n}$, are gaped by metacavity length $L$ along \textbf{x} direction and offset by $\Delta s$ along \textbf{y} direction.  The radii of the cylinders for the two arrays are $\mathbbm{r}_1$ and $\mathbbm{r}_2$, respectively.  The incident wave could be $s$-polarized (electric field along \textbf{z} direction) or $p$-polarized (electric field perpendicular to \textbf{z} direction).}

In this work, we construct most elementary Fabry-P\'{e}rot metacivities based on two parallel metamirrors consisting of a single-layered (one-dimensional, 1D) periodic arrays of circular dielectric cylinders, where constructive interference of multiply reflected beams between the metamirrors leads to discrete Fabry-P\'{e}rot resonances. Based on the explicit formulas for the scattering coefficients of individual cylinders, both reflection efficiency and reflection phase of the metamirror can be analytically calculated through the multiple scattering theory. It is shown that such simple metamirror can provide high reflectivity (up to unit reflection) and broad reflection phase coverage (including the exceptional scenarios of perfect electric and magnetic mirrors).  With the reflection properties of individual metamirrors obtained, as long as the pair of metamirrors are not too close and thus their near-field couplings are negligible, intuitively simple closed-form expressions for the optical responses of the whole Fabry-P\'{e}rot metacavities can be directly derived without any fitting parameters. We further demonstrate, both numerically and analytically, the great flexibilities for tuning reflection properties of metamirrors enable efficient and precise manipulations of the Fabry-P\'{e}rot resonances supported, which can be engineered to resonate over a broad spectral regime with arbitrary Q-factors, including the extreme bound states in the continuum (BICs) with infinite Q-factors.  The capability of achieving ultra-high-Q  Fabry-P\'{e}rot  resonances, and even bound states in the continuum, using simple single-layer dielectric structures suggests a powerful and scalable route toward compact high-performance optical cavities, holding significant potential for a broad range of applications, including low-threshold lasing, enhanced light-matter interactions, and optical sensing.

\section{Theoretical Model and Analytical Formalisms}
\subsection{Two-Dimensional (2D) Scattering by an individual cylinder}

We start with the 2D scattering by an individual cylinder oriented along \textbf{z}-axis (radius $\mathbbm{r}$ and refractive index $\mathbbm{n}$), in the background with $n=1$ (as is the case throughout this study). Here 2D scattering means that the wavevector $\textbf{k}$ of the incident plane wave is on the $\textbf{x}$-$\textbf{y}$ plane ($\textbf{k} \bot \textbf{z}$) and can be $s$-polarized (electric field along $z$ direction: $\mathbf{E_0}\parallel\mathbf{z}$) or $p$-polarized (magnetic field along $z$ direction:  $\mathbf{H_0}\parallel\mathbf{z}$), with angular frequency $\omega$ and free-space wavelength $\lambda$. For an individual cylinder, the 2D scattering efficiency (scattering cross section normalized by the single channel scattering limit $2\lambda/\pi$) is~\cite{Bohren1983_book}:
\begin{equation}
\label{Q_ext}
N_{\rm sca}^{p,s} = \sum\nolimits_{m = -\infty}^\infty|{\mathbbm{a}}^{p,s}_m|^2,
\end{equation}
where $\mathbbm{a}_m^{p,s}$ are the scattering (cylindrical harmonic expansion) coefficients for $p$- and $s$-polarized incident waves respectively, which can be calculated analytically as follows~\cite{Bohren1983_book}:
\begin{eqnarray}
\label{scattering coefficients}
\mathbbm{a}_m^{p}=\mathbbm{a}_{-m}^{p}={{\mathbbm{n}\textbf{J} _{ m} (\mathbbm{n}\alpha )\textbf{J}'_{m} (\alpha ) - \textbf{J}_{m} (\alpha )\textbf{J}'_{m} (\mathbbm{n}\alpha )} \over {\mathbbm{n}\textbf{J} _{ m} (\mathbbm{n}\alpha )\textbf{H} '_{m} (\alpha ) - \textbf{H} _{m} (\alpha )\textbf{J} '_{ m} (\mathbbm{n}\alpha )}},\\
\mathbbm{a}_m^{s}=\mathbbm{a}_{-m}^{s}={{\mathbbm{n}\textbf{J} _{ m} (\alpha )\textbf{J}'_{m} (\mathbbm{n}\alpha ) - \textbf{J}_{m} (\mathbbm{n}\alpha )\textbf{J}'_{m} (\alpha )} \over {\mathbbm{n}\textbf{J} '_{ m} (\mathbbm{n}\alpha )\textbf{H} _{m} (\alpha ) - \textbf{H} '_{m} (\alpha )\textbf{J} _{ m} (\mathbbm{n}\alpha )}}.
\end{eqnarray}
Here $\textbf{J}_m$ and $\textbf{H}_m$ are respectively the first-kind Bessel and Hankel functions of order $m$~\cite{Bohren1983_book}; $\alpha$ is the normalized size parameter $\alpha=k\mathbbm{r}$ ($k=|\mathbf{k}|$);  the accompanying primes indicate their differentiation with respect to the entire argument within the bracket.

\pict[1.1]{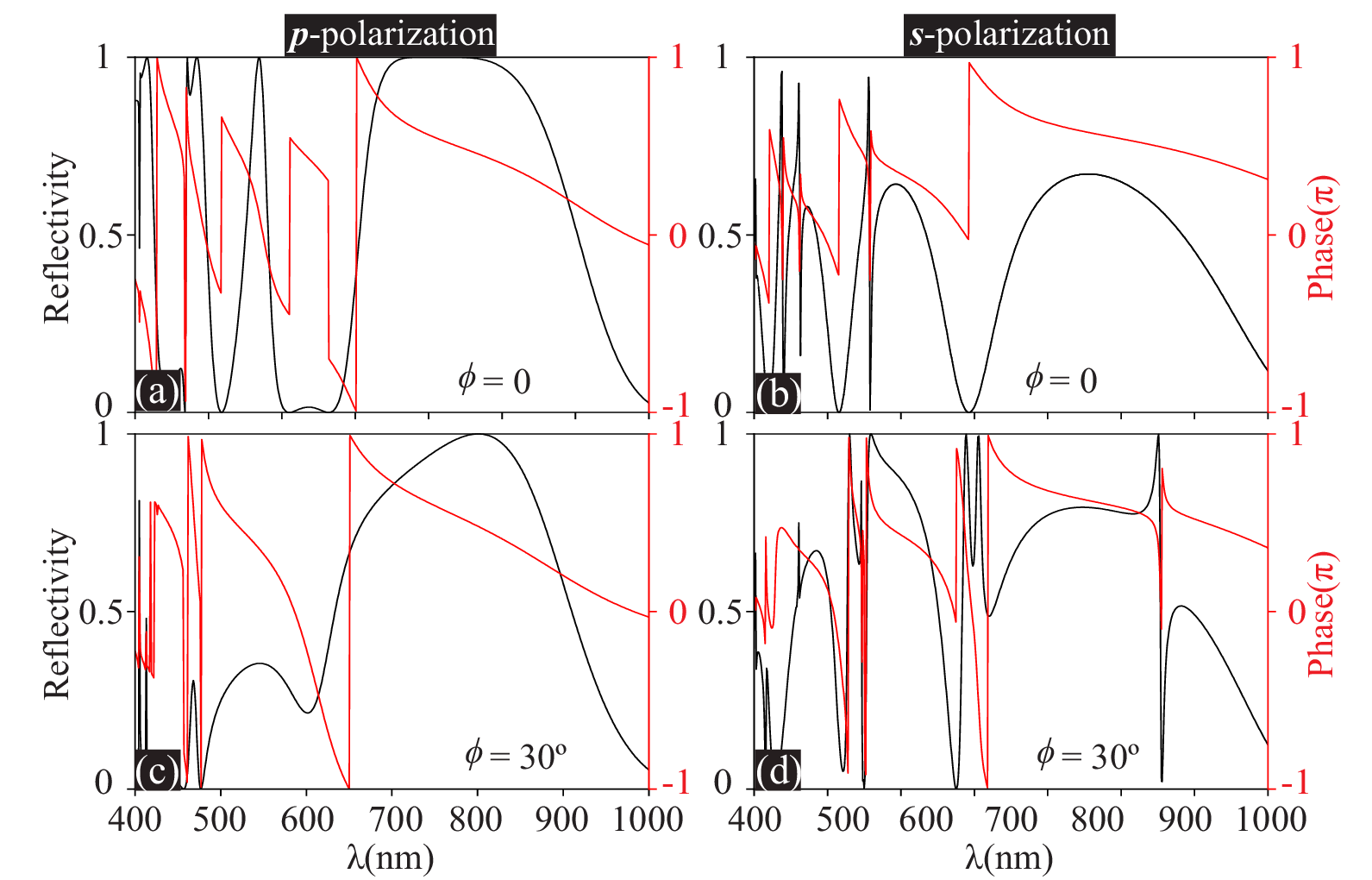}{fig2}{\small  Reflectivity [Eq.~(\ref{reflection_efficiency1})] and reflection phase [Eq.~(\ref{reflection_efficiency2})] spectra (with respect to incident wavelength) for a single-layered array (period $d=280$~nm, refractive index $\mathbbm{n}=3.6$, and cylinder radius $\mathbbm{r}=100$~nm). The incident wave could be $p$-polarized [(a) and (c)] and $s$-polarized [(b) and (d)],  and the results cover both scenarios of normal incidence [$\phi=0$ for (a) and (b)] and oblique incidence [$\phi=30^{\circ}$ for (c) and (d)]. }

\subsection{Optical Properties of a 1D Periodic Cylinder Array}
For a periodic array of cylinders (period is $d$ along \textbf{y}-axis; refer to Fig.~\rpict{fig1}) with incident angle $\phi$, to make sure that the array functions solely as a refelction mirror, we have confined our discussion to the diffractionless regime [$\phi<\arcsin(\lambda/d-1)$] with only zeroth order transmission and reflection. According to the multiple scattering theory~\cite{YASUMOTO_ElectromagneticTheoryandApplicationsforPhotonicCrystals_modeling_2005,liu_generalized_2017}, the cylindrical harmonic expansion coefficients ($a_m^{p,s}$) for any cylinder within the array (the couplings among cylinders have been already taken into consideration) is related to those of an individual cylinder ($\mathbbm{a}_m^{p,s}$) through:
\begin{equation}
\label{lattice_equation}
(\mathbb{I}-\mathbb{T}\cdot\mathbb{C})\textbf{A}=\mathbb{T}\textbf{B},
\end{equation}
where $\mathbb{I}$ is the identity matrix; $\mathbb{T}_{ml}=-\delta_{ml}\mathbbm{a}_m^{p,s}$ ($\delta_{ml}$ is Kronecker delta function; both subscripts $m$ and $l$ denote the order of cylindrical harmonics); $\textbf{A}=a_m^{p,s}$; $\textbf{B}=e^{-im\phi}i^m$; and the coupling effect is embedded into the lattice sum matrix $\mathbb{C}_{ml}$, which can be expressed as~\cite{YASUMOTO_ElectromagneticTheoryandApplicationsforPhotonicCrystals_modeling_2005,liu_generalized_2017}:

\begin{equation}
\label{lattice_matrix}
\mathbb{C}_{ml}=\sum\nolimits_{j = 1}^\infty  {\textbf{H}_{m - l} (jkd)[e^{ik_yj d}  + ( - 1)^{m - l} e^{ - ik_yj d} ]},
\end{equation}
where  $k_y=k\sin{\phi}$ is incident wavevector component along \textbf{y}-axis (projection to the \textbf{x}-axis is $k_x=k\cos{\phi}$) . Through solving Eq.~(\ref{lattice_equation}) with Eq.~(\ref{lattice_matrix}), the expansion coefficients $a_m^{p,s}$ for lattice cylinder can be obtained.   Then for the 1D array (metamirror), the complex zero-order reflection coefficient can be expressed as~\cite{YASUMOTO_ElectromagneticTheoryandApplicationsforPhotonicCrystals_modeling_2005,liu_generalized_2017}:
\begin{equation}
\label{reflection_coefficients}
\mathrm{r}^{p, s}(\phi)= \pm\frac{2}{d k_x} \sum_{m=-\infty}^{\infty}(-1)^me^{im\phi} a_{m}^{p, s},
\end{equation}
where the sign "$\pm$" correspond to incident $p$ and $s$ polarizations, respectively. As a result, the reflection efficiency (reflectivity) $\mathrm{R}^{p, s}(\phi)$ and reflection phase $\varphi^{p, s}(\phi)$ for the metamirror can be directly obtained as follows: 
\begin{eqnarray}
\mathrm{R}^{p, s}(\phi)=|\frac{2}{d k_x} \sum_{m=-\infty}^{\infty}(-1)^me^{im\phi} a_{m}^{p, s}|^2 \label{reflection_efficiency1},\\
\varphi^{p, s}(\phi)=\mathrm{Arg}[\pm\frac{2}{d k_x} \sum_{m=-\infty}^{\infty}(-1)^me^{im\phi} a_{m}^{p, s}] \label{reflection_efficiency2}.
\end{eqnarray}
Here $\varphi=0$ and $\varphi=\pi$ (half-wave loss) correspond to ideal magnetic and electric metamirrors, respectively~\cite{liu_generalized_2017}.

\pict[1.1]{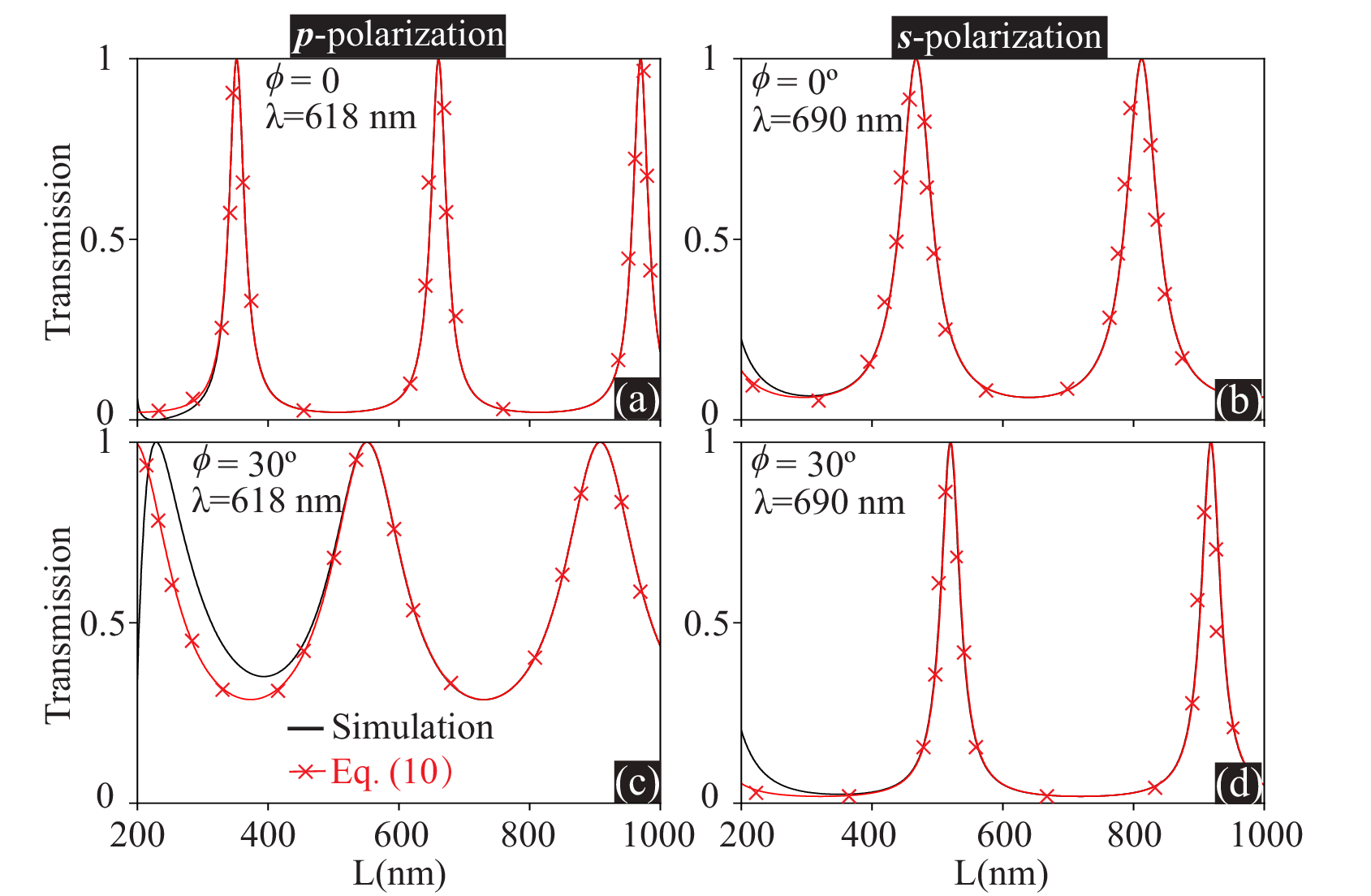}{fig3}{\small  Transmission spectra  with respect to $L$ for the whole metacavity consisting of two identical arrays ($d=280$~nm, $\mathbbm{n}=3.6$, and $\mathbbm{r}_1=\mathbbm{r}_2=100$~nm) without offset ($\Delta s=0$). The incident wave could be $p$-polarized [(a) and (c); $\lambda=618~$nm] and $s$-polarized [(b) and (d); $\lambda=690~$nm], and the two sets of results [numerical simulation through COMSOL and  Eq.~(\ref{transmission-fb-identical})] cover both scenarios of normal incidence [$\phi=0$ for (a) and (b)] and oblique incidence [$\phi=30^{\circ}$ for (c) and (d)]. }

\subsection{Fabry-P\'{e}rot Metacavities with Two Parallel Metamirrors of 1D Dielectric Cylinder Arrays}
\label{section2-3}

We then proceed to Fabry-P\'{e}rot metacavities consisting of two parallel arrays of 1D dielectric cylinders (indicated respectively as metamirrors $\mathds{1}$ and $\mathbbm{2}$, respectively), with the metacavity length $L$ (distance between the two metamirrors), as schematically shown in  Fig.~\rpict{fig1}. For such a metacavity, similar to the single-layered array, its optical responses can indeed be calculated analytically~\cite{YASUMOTO_ElectromagneticTheoryandApplicationsforPhotonicCrystals_modeling_2005}. Nevertheless, the formalisms are too complicated, from  which simple principles and intuitive physical pictures can be barely extracted.  When the near-field couplings between the metamirrors are negligible ($L$ is not too small), the two arrays can be viewed as two independent reflecting mirrors with reflectivities  $\mathrm{R}_{\mathds{1},\mathbbm{2}}$ and reflection phases  $\varphi_{\mathds{1},\mathbbm{2}}$, and then intuitively simple formulas can be derived for the Fabry-P\'{e}rot metacavities to fully capture their optical responses~\cite{YARIV_2006_Photonics}.

In the following discussions, to simplify the expressions, we have dropped the superscripts indicating incident polarizations ($s$ or $p$) and  the bracket containing the incident angle ($\phi$). Both reflectivity ($\mathrm{R}_{\mathds{1},\mathbbm{2}}$) and  reflection phase ($\varphi_{\mathds{1},\mathbbm{2}}$) can be directly calculated according to Eq.~(\ref{reflection_coefficients})-Eq.~(\ref{reflection_efficiency2}). With the reflection properties of each metamirror obtained, the overall transmission efficiency of the whole Fabry-P\'{e}rot metacavity can be expressed as (when the metamirrors are lossless)~\cite{YARIV_2006_Photonics}:
\begin{equation}
\label{transmission-fb}
\rm{T_{\mathbb{FP}}}=\frac{\left(1-R_\mathds{1}\right)\left(1-R_\mathbbm{2}\right)}{\left(1-\sqrt{R_\mathds{1}} \sqrt{R_\mathbbm{2}}\right)^2+4 \sqrt{R_\mathds{1}} \sqrt{R_\mathbbm{2}} \sin ^2(\frac{\delta}{2})},
\end{equation}

\pict[1.1]{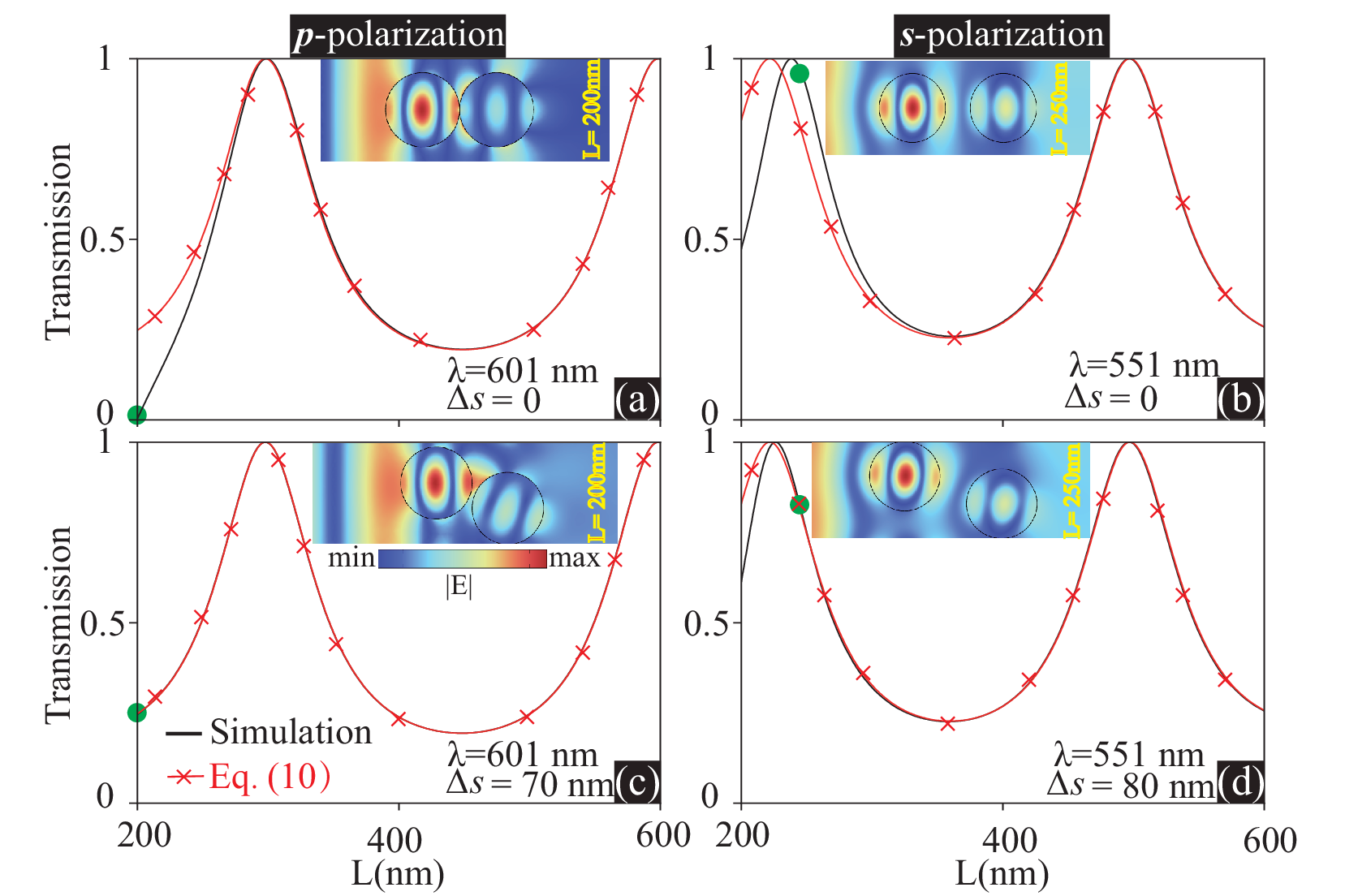}{fig4}{\small  Transmission spectra  with respect to $L$ for the whole metacavity consisting of two identical arrays ($d=280$~nm, $\mathbbm{n}=3.6$, and $\mathbbm{r}_1=\mathbbm{r}_2=100$~nm). The normally incident wave ($\phi=0$) could be $p$-polarized [(a) and (c); $\lambda=601~$nm] and $s$-polarized [(b) and (d); $\lambda=551~$nm], and the two sets of results [numerical simulation and  Eq.~(\ref{transmission-fb-identical})] cover both scenarios with offsets [$\Delta s=70$~nm and  $80$~nm for (c) and (d), respectively] and without offset [$\Delta s=0$ for (a) and (b)]. For all cases, numerical near-field distributions for one unit-cell are included as insets (in terms of electric field amplitude; the corresponding positions are marked by green dots): $L=200$~nm for (a) and (c); $L=250$~nm for (b) and (d).}

where $\delta=2k_xL+\varphi_{\mathds{1}}+\varphi_{\mathbbm{2}}$ is the round-trip phase accumulation between the two metamirrors, considering both the optical-path-length dynamic phase  $2k_xL$ and abrupt phases induced upon reflections $\varphi_{\mathds{1},\mathbbm{2}}$. For the special scenario of two identical metamirrors [$R_\mathds{1}=R_\mathbbm{2}=R_\mathbbm{o}$; $\varphi_{\mathds{1}}=\varphi_{\mathbbm{2}}=\varphi_\mathbbm{o}$; $\delta_\mathbbm{o}=2(k_xL+\varphi_\mathbbm{o})$], Eq.~(\ref{transmission-fb}) would be reduced to the more well-known form:
\begin{equation}
\label{transmission-fb-identical}
\rm{T_{\mathbb{FP}}}=\frac{(1-R_\mathbbm{o})^2}{(1-R_\mathbbm{o})^2+4 R_\mathbbm{o}\sin ^2(\delta_\mathbbm{o} / 2)}.
\end{equation}

For conventionally pedagogical Fabry-P\'{e}rot cavities where  $R$ and $\varphi$ are dispersionless (independent of incident wavelength), Fabry-P\'{e}rot resonances are centered spectrally at discrete equal-spaced $\omega_n$ ($n$ is an integer), where $\rm{T_\mathbb{FP}}$ reaches its maximum value $\frac{\left(1-R_\mathds{1}\right)\left(1-R_\mathbbm{2}\right)}{\left(1-\sqrt{R_\mathds{1}} \sqrt{R_\mathbbm{2}}\right)^2}$ (it reaches ideal unit transmission $\rm{T_{\mathbb{FP}}}=1$ only for two identical mirrors) and $\delta_n=2n\pi$~\cite{YARIV_2006_Photonics,Yang2012_PT,Liu2014_arXiv_Geometric}.

For the general scenarios with non-negligible dispersions, the central positions of Fabry-P\'{e}rot resonances can be located by searching for the poles of $\rm{T_\mathbb{FP}}$ in the complex angular frequency ($\breve{\omega}=\omega^r+i\omega^i$) domain [$\rm{T_\mathbb{FP}}(\breve{\omega})=\infty$]. As a result, the complex resonant frequency is the solution of the following equation:
\begin{align}
\label{complex-resonant}
	&\left(1-\sqrt{R_\mathds{1}(\omega^r)} \sqrt{R_\mathbbm{2}(\omega^r)}\right)^2 \nonumber \\
	&+4 \sqrt{R_\mathds{1}(\omega^r)} \sqrt{R_\mathbbm{2}(\omega^r)} \sin ^2\left[\frac{\delta(\breve{\omega})}{2}\right] = 0,
\end{align}
where the complex phase term $\delta(\breve{\omega})=[2\omega^rL\cos(\phi)/c+\varphi_{\mathds{1}}(\omega^r)+\varphi_{\mathbbm{2}}(\omega^r)]+i[2\omega^iL\cos(\phi)/c]$, and $c$ is the speed of light. The above Eq.~(\ref{complex-resonant})  can be simplified as:
\begin{equation}
\label{complex-resonant-sim}
1-\sqrt{R_\mathds{1}(\omega^r)}\sqrt{R_\mathbbm{2}(\omega^r)}e^{-i\delta(\breve{\omega})}=0.
\end{equation}
To make both real and imaginary parts zero, we have:
\begin{eqnarray}
\label{resonance-1}
2\omega^rL\cos(\phi)/c+\varphi_{\mathds{1}}(\omega^r)+\varphi_{\mathbbm{2}}(\omega^r)=2n\pi,\\
\label{resonance-2}
\exp{[-2\omega^iL\cos(\phi)/c]}=\sqrt{R_\mathds{1}(\omega^r)}\sqrt{R_\mathbbm{2}(\omega^r)}.
\end{eqnarray}
The discrete (indexed by integer $n$) complex resonant frequencies ($\omega_n^r$, $\omega_n^i$) thus obtained can be employed to calculate the Q-factor of the corresponding Fabry-P\'{e}rot resonance:
\begin{equation}
\label{Q-factor}
Q_{n}=\frac{\omega_n^r}{2\omega_n^i}.
\end{equation}

\pict[1.1]{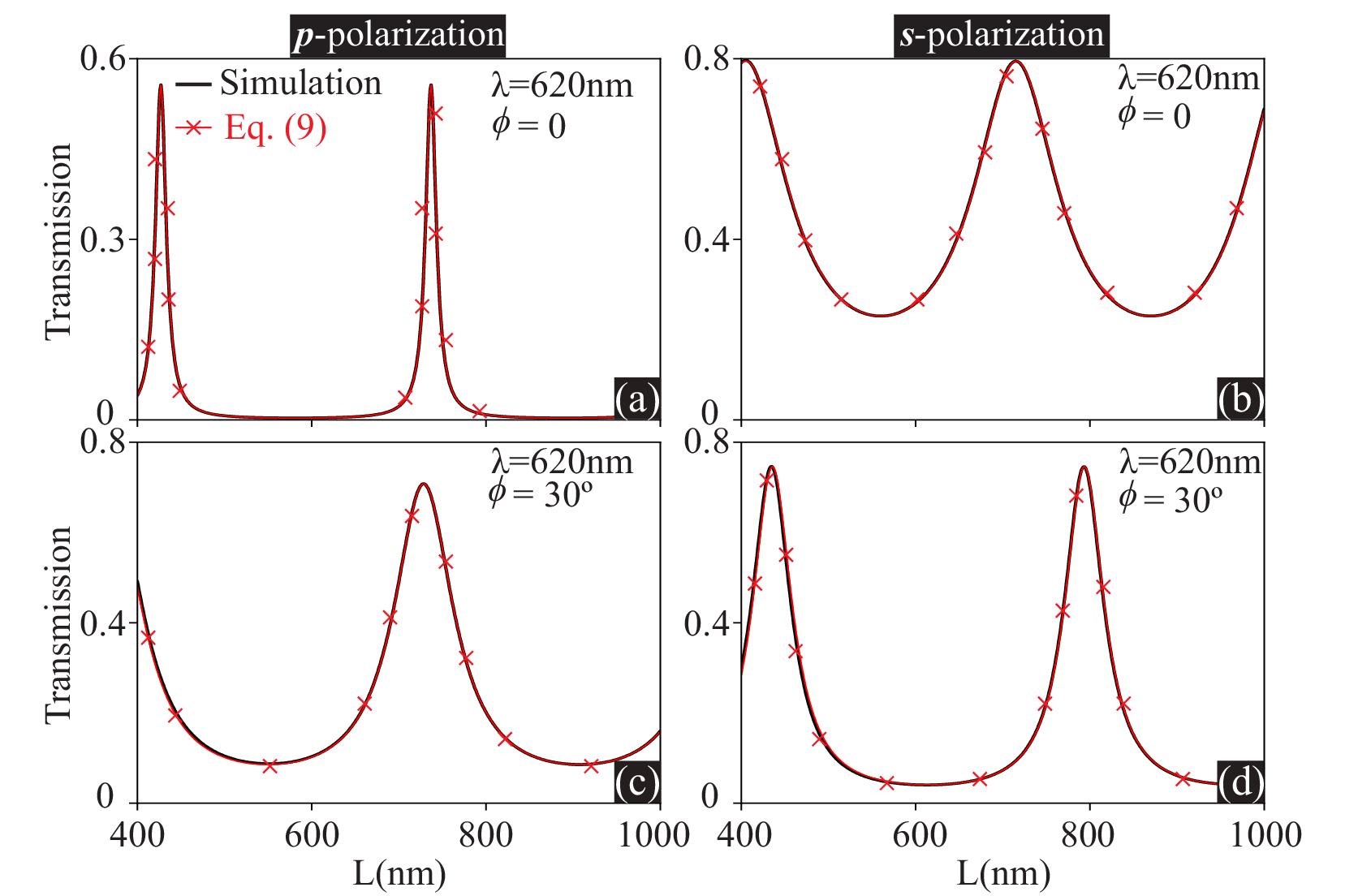}{fig5}{\small  Transmission spectra  with respect to $L$ for the whole metacavity consisting of two different arrays ($d=280$~nm, $\mathbbm{n}=3.6$,  $\mathbbm{r}_1=100$~nm and $\mathbbm{r}_2=75$~nm) without offset ($\Delta s=0$). The incident wave could be $p$-polarized [(a) and (c); $\lambda=620~$nm] and $s$-polarized [(b) and (d); $\lambda=620~$nm], and the two sets of results [numerical simulation and  Eq.~(\ref{transmission-fb})] cover both scenarios of normal incidence [$\phi=0$ for (a) and (b)] and oblique incidence [$\phi=30^{\circ}$ for (c) and (d)]. }

According to Eq.~(\ref{resonance-2}), the larger is the metamirror reflectivity, the smaller is $\omega^i$ and thus the higher is the Q-factor [see Eq.~(\ref{Q-factor})]. The extreme case is perfect reflection for both metamirrors [$R_\mathbbm{1}(\omega^r)=R_\mathbbm{2}(\omega^r)=1$], which results in $\omega^i=0$ and thus $Q=\infty$ of BICs~\cite{HSU_Nat.Rev.Mater._bound_2016, WANG_2024_PhotonicsInsights_Optical}.  Since the reflectivity of metamirrors can be freely tuned ($R\in[0,1]$, as shown in Fig.~\ref{fig2}), covering the ideal perfect mirror of unit reflection $R=1$~\cite{liu_generalized_2017}, according to Eq.~(\ref{resonance-1})-Eq.(\ref{Q-factor}), resonances of arbitrary Q-factors can be predesigned and easily obtained. 

For waves incident at the resonant frequency $\omega=\omega_n^r$, the transmission of the Fabry-P\'{e}rot cavity reaches its local maximum $\rm{T_\mathbb{FP}}(\omega_n^r)=\frac{\left[1-R_\mathds{1}(\omega_n^r)\right]\left[1-R_\mathbbm{2}(\omega_n^r)\right]}{\left[1-\sqrt{R_\mathds{1}(\omega_n^r)} \sqrt{R_\mathbbm{2}(\omega_n^r)}\right]^2}$, which is unit transmission [$\rm{T_{\mathbb{FP}}}(\omega_n^r)=1$] for two identical lossless metamirrors. 

\pict[1.1]{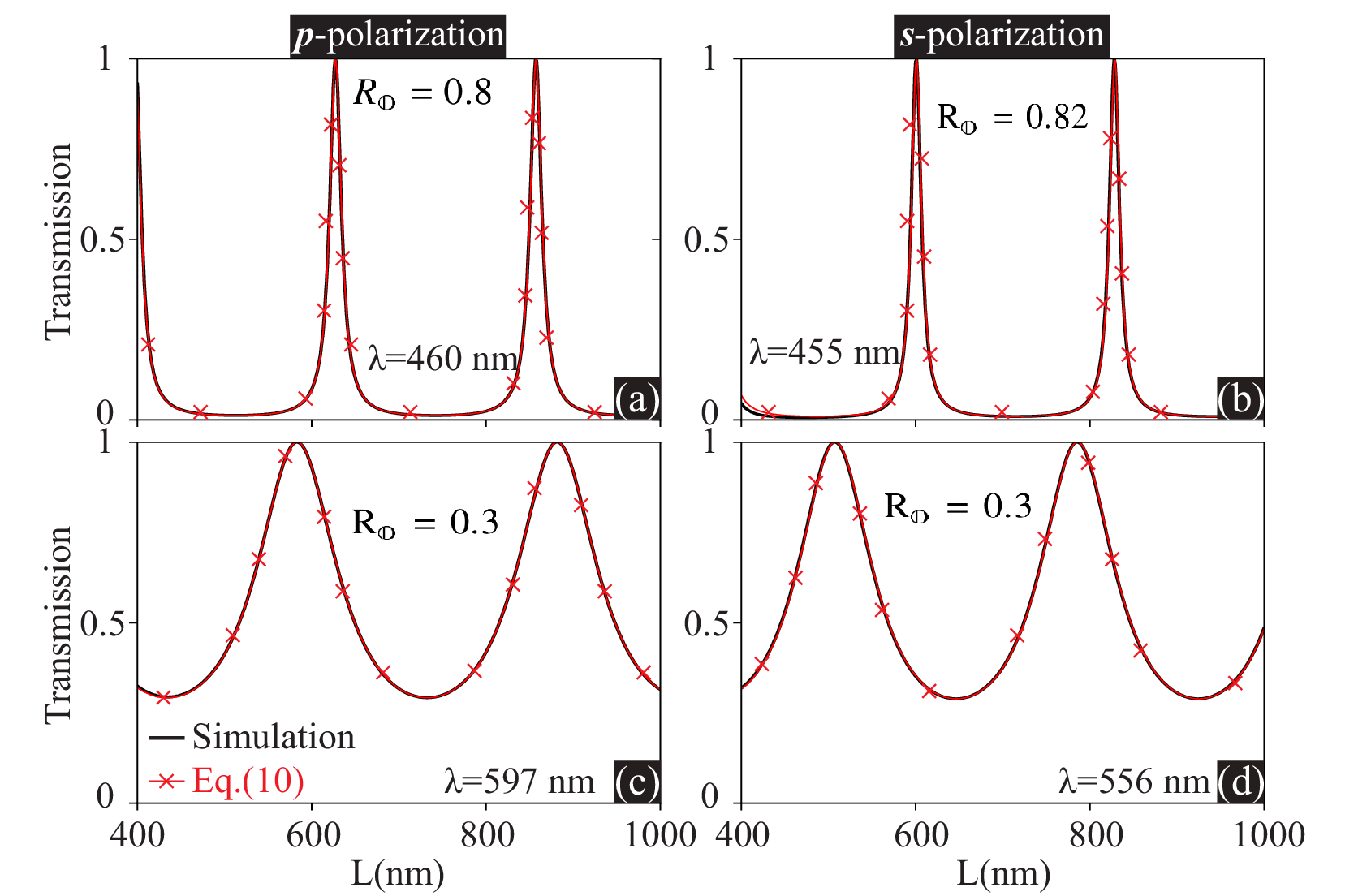}{fig6}{\small Transmission spectra  with respect to $L$ for the whole metacavity consisting of two identical arrays ($d=280$~nm, $\mathbbm{n}=3.6$, and $\mathbbm{r}_1=\mathbbm{r}_2=100$~nm) without offset ($\Delta s=0$). The normally incident wave ($\phi=0$) could be $p$-polarized [(a) and (c)] and $s$-polarized [(b) and (d)], and the two sets of results [numerical simulation and  Eq.~(\ref{transmission-fb-identical})] cover both scenarios with larger  [$R_{\mathbbm{o}}=0.8$ for (a) with fixed $\lambda=460~$nm; ${R}_{\mathbbm{o}}=0.82$ for (b) with fixed $\lambda=455~$nm] and smaller  [${R}_{\mathbbm{o}}=0.3$ for (c) with fixed $\lambda=597~$nm and for (d) with fixed $\lambda=556~$nm] single-layer metamirror reflectivity ${R}_{\mathbbm{0}}$.  }
\section{Results and Discussions}

\subsection{Reflections of Individual Metamirrors}
We start with individual metamirrors, that is single-layered 1D periodic cylinders arrays.  The results [analytically calculated through Eq.~(\ref{Q_ext})-Eq.~(\ref{reflection_efficiency2})] for their reflection properties (in terms of reflectivity and reflection phase) are summarized in Fig.~\ref{fig2}, covering both incident polarizations (\textit{p} and \textit{s}) and both scenarios of normal and oblique incidences. It is clear that the metamirrors provide high reflectivity up to perfect reflection ($R=1$) and a wide range of reflection phase coverage, including perfect electric and magnetic mirrors of respectively $\varphi=\pi$ and $0$~\cite{LIU_2018_Opt.Express_Generalized,liu_generalized_2017}. As a result, those metamirrors render rich degrees of freedom to exploit for flexible manipulations of resonances supported by Fabry-P\'{e}rot metacavities built upon them.

\subsection{Transmissions of Fabry-P\'{e}rot Metacavities}
We then proceed to the simplest  Fabry-P\'{e}rot metacavities consisting of two identical and parallel metamirrors without offset ($\Delta s=0$).  The transmissions of those metacavities are summarized in Fig.~\ref{fig3}, including both normal and oblique incidences for both $p$- and $s$-polarizations.  For each scenario,   the incident wavelength is fixed and two sets of results are shown for the transmission spectra with respect to the cavity length $L$: (i) numerical simulation results obtained through COMSOL Multiphysics; (ii) analytical results based on Eq.~(\ref{transmission-fb-identical}).  For analytical calculations, the parameters in Eq.~(\ref{transmission-fb-identical}) are extracted from individual metamirrors and thus its validity resides on the precondition that the near-field couplings between the two metamirrors are negligible and thus they effectively function independently.  As is clearly shown in Fig.~\ref{fig3}, as $L$ gets smaller and smaller, the discrepancies between the two sets of results become more and more pronounced, indicating the presence of non-negligible near-field couplings.  For two identical metamirrors, the  Fabry-P\'{e}rot resonances are centered at the positions of unit transmission,  which are manifest in Fig.~\ref{fig3}.

\pict[1.1]{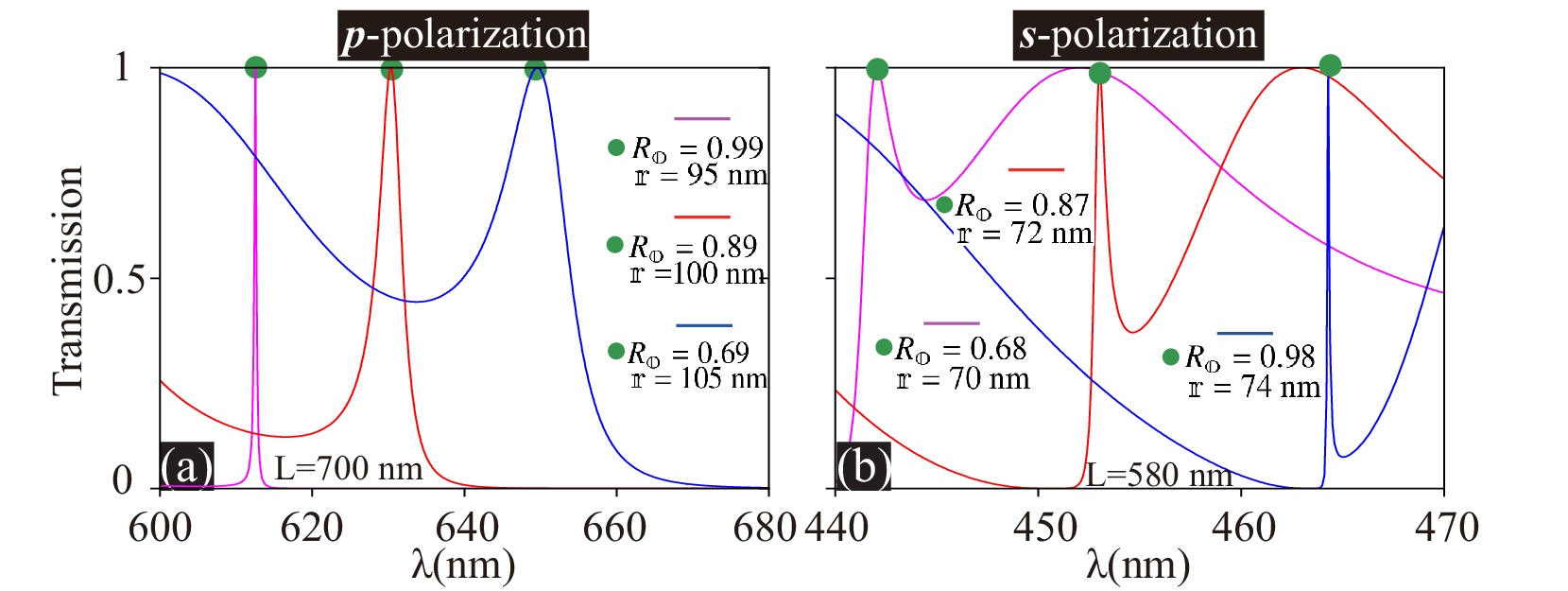}{fig7}{\small Transmission spectra with respect to incident wavelength for the whole metacavity consisting of two identical arrays ($d=280$~nm and $\mathbbm{n}=3.6$) without offset ($\Delta s=0$). The normally incident wave ($\phi=0$) could be $p$-polarized as shown in (a) with fixed $L=700~$nm, and $s$-polarized as shown in (b) with $L=580~$nm. For each polarization, three sets of numerical results with different cylinder radii $\mathbbm{r}$ are showcased: ${R}_{\mathbbm{o}}$ are specified in the figures and they are the single-layer metamirror reflectivity  [Eq.~(\ref{reflection_efficiency1})] at the resonant wavelengths marked by green dots.}

To extend the applicability of our two-independent-metamirror model [Eq.~(\ref{transmission-fb}) and Eq.~(\ref{transmission-fb-identical})] to the small metacavity length regime, we can introduce an offset along \textbf{y} direction ($\Delta s \neq 0$; see Fig.~\ref{fig1}) between the two metamirrors to reduce the inter-metamirror near-field couplings. The comparisons between the scenarios with and without the offsets (with fixed metacavity length) are summarized in Fig.~\ref{fig4}. It is clearly shown that, for both incident polarizations, the introduced offsets have effectively made our model more accurate in the small-$L$ regime, largely due to the suppressed near-field couplings (see the insets of near-field distributions over one unit-cell, where the offsets move the cylinder on the right away from the main field lobe of the first cylinder on the left; the waves are incident from the left).

We have so far discussed only metacavities made of identical metamirrors, and our model is surely applicable to distinct-metamirror metacavity [Eq.~(\ref{transmission-fb})], as confirmed by Fig.~\ref{fig5}. Similar to the identical-metamirror scenario, the Fabry-P\'{e}rot resonances are centered at the positions of round-trip phase accumulation $\delta=2n\pi$. While the difference is that for distinct metamirrors, unit transmission cannot be reached at the resonant positions (as manifest in Fig.~\ref{fig5}), which has already been elaborated on in discussions following Eq.~(\ref{transmission-fb-identical}).

\pict[1.1]{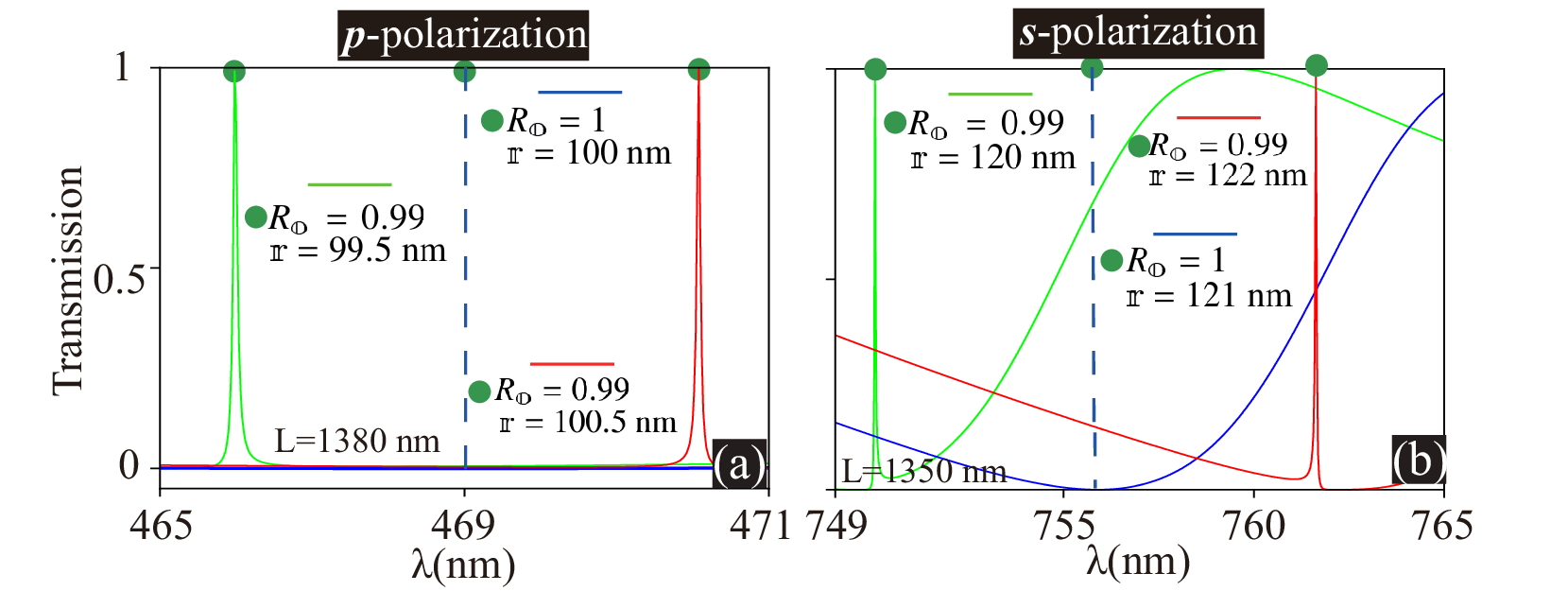}{fig8}{\small The same as that of Fig.~\ref{fig7} except for the parameters specified. The emergence position of infinite Q-factor Fabry-P\'{e}rot resonances (disappearance of resonant transmission peaks) is indicated by the vertical dashed blue line, where the corresponding single-layer metamirror reflectivity is unit (${R}_{\mathbbm{o}}=1$). }

\subsection{Linewidth Reductions for Resonances of Fabry-P\'{e}rot Metacavities}
As has been discussed in Section ~\ref{section2-3}, higher reflectivity of the consisting metamirrors would lead to higher-Q Fabry-P\'{e}rot resonances with reduced  linewidths. This point has been confirmed in both Fig.~\ref{fig6} and Fig.~\ref{fig7}, where we show the transmission spectra with respect to metacavity length and incident wavelength, respectively. For each scenario in Fig.~\ref{fig6}, we have fixed the incident wavelength, and thus also the reflection efficiency and phase of the metamirrors (see Fig.~\ref{fig2}). For both polarizations, at the incident wavelength with higher metamirror reflectivity, the resonance linewidth would be obviously narrower. While for each scenario in  Fig.~\ref{fig7}, the metacavity length is fixed, and the resonance linewidth evolves with varying cylinder radii: the reflectivity specified in the figure corresponds to the metamirror reflectivity (wavelength-dependent; see slo Fig.~\ref{fig2}) at the marked central resonant wavelength; for different cylinder radii, both the resonant wavelength and the corresponding reflectivity of metamirror would vary. As is confirmed by  Fig.~\ref{fig7}, for both incident polarizations, higher resonant reflectivity would result in higher Q-factors and thus also narrower Fabry-P\'{e}rot resonance linewidths. 

The extreme scenario of reduced resonance linewidth is zero linewidth (disappearance of transmission peak) and equivalently infinitely Q-factor Fabry-P\'{e}rot  resonance (BIC in our configuration~\cite{HSU_Nat.Rev.Mater._bound_2016, WANG_2024_PhotonicsInsights_Optical}).  For Fabry-P\'{e}rot cavities, BICs are accessible with only perfect mirrors with unit-reflectivity. This has been demonstrated in Fig.~\ref{fig8}: the dashed blue lines indicate the resonant positions of the BICs ($\delta=2n\pi$) and for the transmission the predicted peaks disappear at those positions. That is, BICs cannot be excited through external sources, as secured by the principle of reciprocity~\cite{CALOZ_2018_Phys.Rev.Applied_Electromagnetic}. As is also clearly shown in Fig.~\ref{fig8}, when the reflectivity deviates from unit, the transmission peaks would emerge, indicating the effective excitation of the un-bounded  Fabry-P\'{e}rot resonances with finite Q-factors. 


\section{Conclusions and Outlook}

We have systematically studied Fabry-P\'{e}rot metacavities formed by two parallel metamirrors consisting of 1D periodic arrays of high-index cylinders. Based on multiple-scattering theory, the reflection amplitude and phase of individual metamirrors are obtained in closed form, enabling  analytical expressions for the transmission and resonance conditions of the resulting metacavities when near-field coupling is negligible. Excellent agreement between analytical predictions and full-wave simulations is achieved. Due to the highly tunable reflection properties of the metamirrors, Fabry-P\'{e}rot resonances can be engineered over a broad spectral range with controllable linewidths and Q-factors. In particular, based on perfect-reflection metamirrors, we have achieved Fabry-P\'{e}rot BICs with infinite Q-factors, and that detailed transitions from leaky resonances to BICs can be fully captured by our developed model.

The  framework (equipped with closed-form expressions) established in this work provides a general and intuitive platform for understanding and designing Fabry-P\'{e}rot resonances in meta-optical systems. It can be readily extended to metamirrors with more complex unit-cell geometries, higher-dimensional periodicities, or dispersion-engineered responses. Incorporating active, nonlinear, or tunable dielectric materials may further enable dynamic control of cavity resonances and Q-factors. The demonstrated ability to realize ultra-high-Q Fabry-P\'{e}rot resonances and even BICs through simple single-layer dielectric structures opens promising opportunities for applications in low-threshold lasing, enhanced light-matter interaction, nonlinear optics, and sensing, positioning dielectric Fabry-P\'{e}rot metacavities as a versatile building block for future nanophotonic devices.

\section*{ACKNOWLEDGMENTS}
This research was funded by the National Natural Science
Foundation of China (12574352)
and projects of Hunan Province (2024JJ2056).\\

\section*{Data availability}
The data that support the findings of this article are openly
available.


\end{document}